\documentclass[12pt]{article}
\usepackage{amsmath,amsfonts,amssymb}
\usepackage{graphicx}
\usepackage{psfrag}

\numberwithin{equation}{section}

\begin{document}
\newcommand{\todo}[1]{{\em \small {#1}}\marginpar{$\Longleftarrow$}}   
\newcommand{\labell}[1]{\label{#1}\qquad_{#1}} 
\newcommand{\ud}{\mathrm{d}}

\rightline{DCPT-05/39}   
\vskip 1cm 

\begin{center} 
{\Large \bf Winding tachyons in asymptotically supersymmetric black strings}
\end{center} 
\vskip 1cm   
  
\renewcommand{\thefootnote}{\fnsymbol{footnote}}   
\centerline{\bf Simon 
F. Ross\footnote{S.F.Ross@durham.ac.uk}}     
\vskip .5cm   
\centerline{ \it Centre for Particle Theory, Department of  
Mathematical Sciences}   
\centerline{\it University of Durham, South Road, Durham DH1 3LE, U.K.}   
  
\setcounter{footnote}{0}   
\renewcommand{\thefootnote}{\arabic{footnote}}

\vskip 1cm
\begin{abstract}
  We show that in the D1-D5 system with angular momentum, there can be
  localised tachyonic winding string modes in the interior of the
  spacetime even if we choose a spin structure which preserves
  supersymmetry in the asymptotic region. We consider cases where the
  tachyonic region extends outside the event horizon, and argue that
  the natural endstate of tachyon condensation in almost all cases is
  one of the solitonic solutions which correspond to special
  microstates of the D1-D5 system.
\end{abstract}

\section{Introduction and Summary}

The condensation of localised closed string tachyons has been found to
have a rich physics.\footnote{Useful reviews of these
  developments, with more extensive references,
  are~\cite{diss,over}.} In the pioneering work of~\cite{panic}, it was
argued that the condensation of a twisted sector tachyon could cause
an orbifold singularity to decay into flat space. This was generalised
to twisted circles in~\cite{twist}, and an interesting discussion of the null
orbifold was recently given in~\cite{bk}.  The condensation of a
winding tachyon in a spacetime with a circle with antiperiodic
fermions has been argued to lead to the pinching off of cycles,
changing the spatial topology~\cite{apart}, and to the decay of a
black string into a static or expanding bubble~\cite{hor:tachyon}. It
has been conjectured that such winding string tachyon condensation can
provide a chronology protection mechanism~\cite{chron}. The
condensation of a winding tachyon on a time-dependent circle has even
been argued to provide a perturbative resolution of big-bang like
singularities~\cite{silv:tachyon}.

The aim of this paper is to provide a further interesting example of
tachyon condensation, modifying the example discussed
in~\cite{hor:tachyon} so that the spacetime is asymptotically
supersymmetric,  so that the only instability is the
localised winding tachyon. 

In~\cite{hor:tachyon}, a black string wrapped on a circle with
antiperiodic boundary conditions was considered. The circle will
shrink near the string, so for a suitable choice of parameters there
is a winding mode which becomes tachyonic outside the event horizon.
It was argued that the condensation of this tachyon will cause the
black string to decay into a charged bubble solution with no horizon.

This example provides a new insight into the endpoint of Hawking
evaporation for such black strings. However, the antiperiodic boundary
conditions give a non-zero Casimir energy density in the asymptotic
region. This will cause the circle to contract, and its back-reaction
could spoil the asymptotic flatness of the solution. Also, flat space
with this supersymmetry-breaking compactification has a
non-perturbative instability to decay to the Witten `bubble of
nothing'~\cite{witten:bubble}. Although the tachyon condensation is
more dynamically efficient than this non-perturbative decay, this does
show that the instability of the black string solution is not truly
localised. In~\cite{hor:tachyon}, it was suggested that these issues
could be addressed by taking the circle to be a cycle in a Riemann
surface of the type considered in~\cite{eva:handle}, where fluxes and
branes would stabilise the circle. They nonetheless imply it is worth
seeking an example of a black string solution with a similar winding
string tachyon, but where the circle has periodic boundary conditions
in the asymptotic region, so that supersymmetry will be restored
asymptotically, and the tachyon provides the only instability.

This sounds like an impossible request, since to have a tachyon in the
winding sector, the fermions have to be antiperiodic around a
circle. However, as we will see below, it can actually be easily
achieved by exploiting the idea underlying the smooth solitons found
in~\cite{gadsme,gadsmald,lunin,gms1,gms2,nonbps}: in the D1-D5 system with
angular momentum, in the asymptotically supersymmetric sector, we can
twist the asymptotic $S^1$ by a rotation in the $S^3$ to obtain a
circle with antiperiodic fermions in the near-horizon region.  Such a
circle could then have a tachyon in a string winding sector.

We develop this example in the next section, and discuss its
near-horizon limit in section~\ref{nh}. We will see that as in the
case without angular momentum, there are choices of parameters in the
D1-D5 system for which a tachyon appears outside the event horizon,
and the curvatures and string coupling are small at the horizon.
Physically what is happening is that the interaction between the
charges and the angular momentum can be tuned so that the black string
causes some twisted circle to shrink.  The near-horizon limit of these
solutions is found to be identical to the case with zero angular
momentum, clearly showing that the local analysis of the tachyon near
the horizon for these examples will be exactly the same despite the
change in asymptotic spin structure.

It is important to note that the appearance of a tachyon on a twisted
circle is very special to the example we have considered. It is not
simply a matter of adding angular momentum; the twisted circle has
small size near the horizon as a result of the presence of a $dy
d\phi$ cross term in the metric, and such cross terms appear to be
special to the D1-D5 system.\footnote{The other special feature of the
  D1-D5 system is that this is the only case with a regular
  supersymmetric rotating black hole solution~\cite{bmpv} (see
  \cite{Gauntlett,herdeiro} for further discussion of this fact).  One
  wonders if these facts are in any way related. } The cross term
disappears if we set one of the charges to zero, and no such term
appears in the metric for a rotating $p$-brane with a single
charge~\cite{cyint,klt,har}. The presence of the cross term is also
(obviously) not duality-invariant, although the tachyon should appear
in some other way in related duality frames (see discussion below).

In section~\ref{endp}, we consider the endpoint of this tachyon
condensation. Here we find a major difference relative to the previous
case: In~\cite{hor:tachyon}, there was a static bubble only for
certain values of the parameters, and the black string otherwise
decayed into an expanding bubble which eventually consumed the entire
spacetime. By contrast, we find that there is a smooth stationary
soliton solution from~\cite{nonbps} for all values of the parameters.
This provides a natural endpoint for the decay of the tachyon in
almost all cases. The only exception is the case of unit twist, where
the soliton is BPS, and turns out to be larger than the tachyonic
black string if the charge is large, $Q/R^2 \gg1$. It is not clear
what happens in this case: the decay may eventually settle down to the
soliton after some non-trivial evolution, or there may some other
endstate for the decay, although no such solution is explicitly
identified.  We argue that expanding bubbles do not appear in the
decay of these black strings.  It would be valuable to have a deeper
understanding of the connection between this result and the choice of
spin structure.

In addition to providing a new example of localised closed string
tachyon condensation, this analysis gives an intriguing new relation
between the black string solutions and the smooth solitons. Mathur and
collaborators have argued that these solitons provide a geometrical
description of the microstates responsible for the black string's
entropy, and that the black string only emerges as a `coarse-grained'
description of the microstate geometries \cite{lm2,stat} (see
\cite{mathurrev} for a review). It is therefore quite interesting that
this tachyon condensation provides a stringy mechanism by which a
`na\"\i ve' black string geometry can decay into one of the smooth
solitons. However, there are plenty of black strings which do not have
such an instability: our argument only applies to the special class of
solutions where a circle is smaller than the string scale outside of
the horizon. Thus, while this observation is suggestive, its relation
to the Mathur programme is not entirely clear.

In~\cite{hor:tachyon}, it was argued that any black string will
eventually decay by Hawking radiation until it reaches the point where
this instability sets in. The decay to a soliton does not seem to have
the same inevitability in our case, since the black strings which
decay via the tachyon condensation have a non-zero angular momentum,
and we would expect that any initial angular momentum on a macroscopic
black string will quickly be lost through preferential Hawking
radiation of co-rotating modes (although a detailed analysis of this
has not been carried out for these higher-dimensional black holes).
That is, a generic black string will shrink by Hawking radiation, but
it may not reach a regime where a circle with antiperiodic boundary
conditions is string scale outside the horizon before the curvatures
reach the string scale, and corrections to the classical geometry
become important.

Finally, we can study this system when the asymptotic circle is
smaller than the string scale, which will make a prediction for the
behaviour of the T-dual spacetime. The periodic spin structure implies
the spacetime is still stable asymptotically, and our analysis still
predicts a winding mode tachyon in the region near the horizon for
appropriate choices of the parameters. In the T-dual system, this
winding string gets mapped to a string with momentum along a dual
circle, wrapped on an $S^1$ in the $S^3$. Although this is a
contractible circle, which does not correspond to a conserved winding
number, wrapping this circle may play a role in making this state
tachyonic, via a coupling similar to that of long strings~\cite{ls}. A
project for future investigation is to see if we can identify and
understand this localised tachyon in the dual system.\footnote{An
  effective action with a field which becomes tachyonic near the event
  horizon of a charged black hole was recently discussed
  in~\cite{Friess}. This might be a useful model for the T-dual
  physics.}  The simplest case to consider is probably the NS5-P
system.  We would expect that the soliton solutions would still be
appropriate endpoints of tachyon condensation, even though they are
not smooth spacetimes in the T-dual variables. They would be
interpreted as geometries with explicit brane sources.

\section{Black strings with periodic boundary conditions}

The string-frame metric for the D1-D5 solution with one angular
momentum is ~\cite{cvetic,cvetlars} (in a form previously employed
in~\cite{nonbps})
\begin{eqnarray} \label{2chmetric}
  \mathrm{d}s^2&=&\frac{1}{\sqrt{\tilde{H}_{1}\tilde{H}_{5}}}
  \left[-(f-M)(\mathrm{d}t- (f-M)^{-1}M \cosh\delta_1 \cosh \delta_5
    a_1 \cos^2\theta \mathrm{d}\psi)^2\right.
  \\&&+\left.f(\mathrm{d}y \nonumber +f^{-1}M \sinh \delta_1 \sinh
    \delta_5 a_1 \sin^2\theta \mathrm{d}\phi)^2 \right]+
  \sqrt{\frac{\tilde{H}_1}{\tilde{H}_5}}\sum_{i=1}^4 \mathrm{d}z_i^2
  \\
  &&+\sqrt{\tilde{H}_{1}\tilde{H}_{5}}\left(\frac{\mathrm{d}r^2}{r^2+a_{1}^2-M}
    +\mathrm{d}\theta^2+\frac{r^2\sin^2\theta}{f}\mathrm{d}\phi^2
    +\frac{(r^2+a_1^2-M)\cos^2\theta}{f-M}\mathrm{d}\psi^2\right) .
  \nonumber
\end{eqnarray}
We are focusing on the simplest case where the twisting phenomenon
occurs, two charges and a single angular momentum. It is useful to
focus on the case with no momentum along the $y$ circle, as it makes
it clear such a momentum is not playing a role in this story, but the
discussion could easily be generalised to include momentum charge and
two angular momenta.  The dilaton field is $e^{2\Phi} =
\tilde{H}_1/\tilde{H}_5$.  See~\cite{gms1} for the value of the RR
field, which we will not need.  The functions appearing in the above
metric are
\begin{eqnarray} 
\tilde{H}_{i}=f+M\sinh^2\delta_i, \quad
f=r^2+a_1^2\sin^2\theta.
\end{eqnarray}

The $z_i$ are coordinates on a $T^4$ of volume $V$, and we take $y$ to
be a periodic coordinate on an $S^1$, $y \sim y+2\pi R$.  In units
where $G^{(5)} = G^{(10)}/(2\pi RV) = \pi/4$, the mass, angular
momentum and charges are
\begin{equation}
M_{ADM} = \frac{M}{2} (\cosh 2\delta_1 + \cosh 2\delta_5 + 1),
\end{equation}
\begin{equation} \label{jpsi}
J_\psi = -  M a_1 \cosh \delta_1 \cosh \delta_5,  
\end{equation}
\begin{equation} \label{c1}
Q_1 = M \sinh \delta_1 \cosh \delta_1,
\end{equation}
\begin{equation}
Q_5 = M \sinh \delta_5 \cosh \delta_5.
\end{equation}
If $Q_1=Q_5$, the dilaton is constant, and the S-dual of this solution 
is an F1-NS5 system with the same metric \eqref{2chmetric}. This is then
the generalisation of the two-charge black string considered
in~\cite{hor:tachyon} to include angular momentum. 

This solution has a horizon at $r_+^2 = M - a_1^2$. Assume $M >
a_1^2$, so that we are discussing a \textit{black} string, with a
regular event horizon. The $y$ circle is then not contractible in the
interior of the spacetime, so we are free to choose the spin structure
on this circle. We will take periodic boundary conditions for the
fermions on the $y$ circle, so that in the asymptotic region $r \to
\infty$, the solution approaches $\mathbb{R}^{5,1} \times T^4 \times
S^1$ with unbroken supersymmetry. 

We then need to identify a suitable circle with antiperiodic fermions.
The Killing vector
\begin{equation}
\xi = \partial_y - \frac{m}{R} \partial_\phi
\end{equation}
has closed orbits for integer $m$. If the integer $m$ is related to
the parameters of the solution by
\begin{equation} \label{mc}
m = \frac{R a_1}{M \sinh\delta_1 \sinh \delta_5}, 
\end{equation}
the proper size of this $S^1$ vanishes as $r^2 \to 0$. Thus, this
circle shrinks to zero size in the interior of the spacetime. If $m$
is odd, the fermions are antiperiodic on this circle, and we expect
that the spectrum of winding modes around this circle includes a
tachyon.\footnote{In fact, such a tachyon should exist for some open
  range of parameters around the values implied by \eqref{mc}, for
  which this circle becomes smaller than the string scale for $r^2$
  near zero. We focus however on the specific choice of parameters
  in~\eqref{mc}, as this will allow us to make more definite
  statements.} In the case without angular momentum, we know that the
tension of the wrapped black string will make the asymptotic circle
shrink near the string. Here, the combination of this effect with the
$dy d\phi$ cross term in the metric causes a different, twisted circle
to shrink when we tune the angular momentum as in~\eqref{mc}. Note
that the asymptotic circle remains of finite size at $r^2=0$ for any
nonzero $a_1$, even though we have not introduced any momentum along
this direction.\footnote{The $y$ circle does go to zero size for
  $r^2=0, \theta=0$, but this is special to the case with one angular
  momentum. In the general case, it would remain of finite size for
  all $\theta$.}

For $M > a_1^2$, $r=0$ is a spacelike surface, so this is similar to
the end of the universe tachyon studied in~\cite{silv:tachyon}. It
would be interesting to further explore the time-dependent physics of
the tachyon condensation behind the horizon of the black string. In
this paper, however, we content ourselves with studying the
generalisation of~\cite{hor:tachyon}, the case where the region in
which a winding mode is tachyonic extends outside the event horizon.
The proper size of this twisted circle at the horizon is given by
\begin{equation} \label{circle}
||\xi||^2 R^2 = \frac{r_+^2 R^2}{\sqrt{\tilde{H}_1 \tilde{H}_5}} \left[ 1 +
    \frac{a_1^2 \sin^2 \theta}{M^2 \sinh^2 \delta_1 \sinh^2 \delta_5} (f +
    M \sinh^2 \delta_1+ M \sinh^2 \delta_5 ) \right]. 
\end{equation}
Thus, for sufficiently small values of $r_+^2$, the proper size of the
orbits of $\xi$ is smaller than the string scale outside the event
horizon.  

These geometries thus provide examples where we expect tachyon
condensation to change the geometry outside the event horizon, and the
geometry at large distances is supersymmetry-preserving, so the
instability is, at least initially, a genuinely localised phenomenon.

\section{Near-horizon limit}
\label{nh}

In the previous section, we argued heuristically that there should be
a winding string tachyon where the twisted circle is small enough.
However, there is only one conserved winding number in the spacetime,
not one for every choice of $m$: the circle we are considering can be
smoothly deformed into the $y$ circle (or any other such twisted
circle) at constant $r$. One might therefore worry that a careful
analysis of the spectrum might find that the supersymmetry on the $y$
circle forbids the appearance of tachyons.  In this section, we
consider the near-horizon limit of \eqref{2chmetric}, where it is
easier to work out the spectrum of string states, and can see that
there really is a tachyon. This will also allow us to relate this case
to the case studied in~\cite{hor:tachyon}.

If we consider near-extremal black strings with the radius $R$ large
compared to the other scales in the geometry, $R \gg \sqrt{Q_1},
\sqrt{Q_5}, \sqrt{M}$, the geometry \eqref{2chmetric} has a near
horizon limit which is locally AdS$_3 \times S^3$ with radii
$\ell_{AdS} = \ell_{S} = \ell = (Q_1 Q_5)^{1/4}$. For the black string
solution, the near-horizon geometry is a BTZ black hole $\times
S^3$~\cite{clads},
\begin{eqnarray}
  \mathrm{d}s^2 &=& - \left(\frac{ \rho^2}{\ell^2} - M_3  \right)
  d\tau^2 + \left(\frac{ \rho^2}{\ell^2} 
    - M_3  \right)^{-1} d\rho^2 + \rho^2 
  d\varphi^2  \nonumber  \\ && + \ell^2 \left[ d\theta^2
  +
  \sin^2 \theta \left( d\phi + m d\varphi \right)^2
  + \cos^2 \theta \left(d \psi + \frac{m}{\ell}
   d \tau\right)^2 \right],\label{aads}
\end{eqnarray}
with 
\begin{equation}
M_3 = \frac{R^2}{\ell^4} (M-a_1^2),
\end{equation}
where we have defined the new coordinates 
\begin{equation}
\varphi = \frac{y}{R}, \quad \tau = \frac{t \ell}{R} \quad \rho^2 =
\frac{R^2}{\ell^2} r^2. 
\end{equation}

By assumption, the fermions are periodic around the $\varphi$ circle.
However, if we make the large coordinate transformation $\tilde{\phi}
= \phi + m \varphi$, $\tilde \psi = \psi + m \tau/\ell$ to bring the
metric to a direct product form, the fermions will be antiperiodic
around $\varphi$ at fixed $\tilde{\phi}, \tilde \psi$ if $m$ is odd.
This solution then agrees exactly with the near-horizon limit of the
two-charged black strings without angular momentum considered
in~\cite{hor:tachyon}. This direct product geometry is just a quotient
AdS$_3 / \mathbb{Z} \times S^3$, and using the methods of
~\cite{mm1,mm2} one can explicitly check that there is a tachyon in
the twisted sector of this orbifold when the circle becomes smaller
than the string scale. This near-horizon limit also shows that the
curvatures and string coupling can be kept small at the horizon, so
that corrections to this perturbative worldsheet analysis are under
control.\footnote{Also, the radial proper distance over which the
  circle remains small is of order the AdS scale $(Q_1 Q_5)^{1/4}$, so
  the circle's size is varying slowly for a large AdS space.}

This system clearly deserves further study. Considering the geometry
with NS fluxes, it ought to be possible to study this tachyon from the
worldsheet point of view, following~\cite{panic,apart,bk}.  For now,
the lesson to draw from this is that in the near-horizon region, the
tachyon story is the same, independent of whether the compact circle
direction in the full asymptotically flat solution has periodic or
anti-periodic fermions.  Thus, we believe there is a winding string
tachyon in the full asymptotically flat solution considered in the
previous section.

\section{Endpoint of tachyon condensation}
\label{endp}

Having argued that there is a tachyon for some black string solutions
of the form~\eqref{2chmetric}, we should now investigate what the
endstate of condensation of this tachyon is. 

In the near-horizon limit discussed in the previous section, the
natural candidate for the endpoint is global AdS$_3 \times S^3$. This
immediately suggests that for the near-extremal black strings which
have this near-horizon limit, the natural endpoint for the decay of
the tachyon will be the soliton of~\cite{nonbps} which embeds global
AdS$_3 \times S^3$ in an asymptotically flat spacetime, with the $S^1$
direction in AdS$_3$ corresponding to the circle labelled by the same
Killing vector $\xi =
\partial_y - \frac{m}{R} \partial_\phi$.\footnote{Note if the
  curvature of the metric near the horizon is small, we can only make
  the circle associated with one choice of $m$ small.} But what about
more general black strings: do they also decay into smooth solitons?

Here there is a significant difference between the present discussion
and~\cite{hor:tachyon}. It was observed in~\cite{hor:tachyon} that the
static bubble solution only existed for a range of values of the
parameters $Q, R$. In our case, however, it was found in~\cite{nonbps}
that there is a soliton solution for any choice of $Q_1, Q_5$, and
$R$ for each positive integer $m$. If we consider for simplicity $Q_1
=Q_5 = Q$, we can illustrate this explicitly: the radius of the circle for
a smooth soliton is fixed by
\begin{equation} \label{rc}
R = \frac{\tilde{M} \sinh^2 \tilde \delta}{\sqrt{ \tilde{a}_1^2 - \tilde{M}}},
\end{equation}
where $\tilde{M}, \tilde{a}_1$ are the parameters for the smooth
soliton, which need not agree with the parameters of the original
black string solution. Together with \eqref{mc}, this implies
\begin{equation} 
\tilde{M} = \frac{\tilde{a}_1^2}{m^2}(m^2-1). 
\end{equation}
The conditions (\ref{mc},\ref{rc}) and the expression for the charge,
$Q = \tilde M \sinh \tilde \delta \cosh \tilde \delta$, fix the
parameters $\tilde M, \tilde a_1$ and $\tilde \delta$. We use
\eqref{mc} and \eqref{rc} to write $\tilde a_1$ and $\tilde M$ in
terms of $\tilde \delta$.  The expression for the charge then yields
an equation for $\tilde \delta$,
\begin{equation} \label{soleq}
\frac{Q}{R^2} (m^2-1) \tanh^3 \tilde \delta + \tanh^2 \tilde \delta -
1 =0. 
\end{equation}
For $m=1$, the solution is $\tanh \tilde \delta =1$. For $m\neq 1$, we can
rewrite this as an expression for $Q/R^2$ in terms of $\tilde \delta$:
\begin{equation} \label{soldelta}
\frac{Q}{R^2} = \frac{\cosh \tilde \delta}{(m^2-1) \sinh^3 \tilde \delta }. 
\end{equation}
The RHS is unbounded, so in contrast to~\cite{hor:tachyon}, a smooth
soliton solution exists for any $R, Q$. (Note also that this is a
monotonic function of $\tilde \delta$, so there is a single soliton solution
for each $Q, R$.) Thus, we can conjecture that whenever the
asymptotically supersymmetric black string solution~\eqref{2chmetric}
has a winding string tachyon, the endpoint of tachyon condensation is
a smooth soliton solution.

If the black string solution is going to decay into the soliton
solution, it must also have a larger mass. In the near-horizon limit,
the BTZ black hole clearly has larger energy than global AdS, so this
condition is satisfied. As we now see, this remains true for the
asymptotically flat solution, at least when it satisfies~\eqref{mc}.

The condition~\eqref{mc} and the condition that we fix the charge
leave us one free parameter in the black string, but it can only vary
in a small range: the black string solution only has a tachyon outside
the horizon if $M-a_1^2$ is small (and positive). The minimum value of
the mass for fixed charge is attained when we have the smallest
possible value for $M$, so we will take
\begin{equation}
M = a_1^2
\end{equation}
in considering the black string. We can then solve for the parameters
in terms of the charges, which gives a very similar expression for
$Q/R^2$, 
\begin{equation} \label{bseq}
\frac{Q}{R^2} m^2 \tanh^3  \delta + \tanh^2 \delta -
1 =0, 
\end{equation}
or
\begin{equation}
\frac{Q}{R^2} = \frac{\cosh \delta}{m^2 \sinh^3 \delta }.
\end{equation}
The small change in the denominator on the RHS implies that that for a
given value of $Q/R^2$, $\delta < \tilde \delta$ (also true for $m=1$,
as $\tilde \delta = \infty$) so for fixed charge $M > \tilde M$, and
hence the ADM mass of the black string is always larger than the ADM
mass of the soliton with the same values of $Q, R$ and $m$.

In~\cite{hor:tachyon}, another condition was imposed: the size of the
sphere at the static bubble was required to agree with the size of the
sphere at the horizon in the original black string. This is a
reasonable condition, corresponding to the intuition that the tachyon
condensation should be happening locally near the horizon. Imposing it
would seem likely to cause problems, however, since the soliton
geometry is fixed by the choice of $Q$ and $R$, so we have no free
parameters left to tune. Surprisingly, we will see that this condition
is also satisfied in almost all cases.

As a warm-up, we discuss the two charge case in~\cite{hor:tachyon}
in a way that makes it easy to compare to our case. The black string
is parametrised by two parameters $r_0, \alpha$, in terms of which the
charge is
\begin{equation} \label{hq}
Q = r_0^2 \sinh 2\alpha.
\end{equation}
To have a tachyon outside the horizon, we choose the asymptotic circle
periodicity to be
\begin{equation} \label{hl}
R = l_s \cosh \alpha,
\end{equation}
which implies $\alpha \gg 1$ so that $R \gg l_s$. The static bubble
solution obtained by analytic continuation has parameters $\tilde r_0,
\tilde \alpha$, and is smooth if
\begin{equation}
R = 2\pi \tilde r_0 \cosh^2 \tilde \alpha.
\end{equation}
The charge $Q$ is given by the same expression \eqref{hq}, so $\tilde
\alpha$ is determined by  
\begin{equation} \label{horalpha}
\frac{Q}{R^2} = \frac{\sinh \tilde \alpha}{\pi \cosh^3 \tilde
  \alpha} 
\end{equation}
(compare \eqref{soldelta}). This has a maximum as a function of
$\tilde \alpha$ at $\tilde \alpha \approx 1$. Thus, for $Q/R^2 \lesssim
1$, there are two soliton solutions. We will be interested in the one
with $\tilde \alpha \geq 1$. The charge is the same, so the ratio of
proper sizes of the sphere at the horizon or bubble is
\begin{equation}
\frac{r_0^2 \cosh^2 \alpha}{\tilde r_0^2 \cosh^2 \tilde \alpha} =
\frac{\tanh \tilde \alpha}{\tanh \alpha}.
\end{equation}
Now $\alpha \gg 1$, and for $Q/R^2 \lesssim 1$, we can solve
\eqref{horalpha} for $\tilde \alpha \geq 1$. So for $Q/R^2 \lesssim 1$,
there is a static bubble which is roughly the same size as the black
string.\footnote{In~\cite{hor:tachyon}, this result was obtained by
  considering general initial data, and finding that when $Q/R^2 \lesssim
  1$, the bubble which was the same size as the black string horizon
  was a local minimum of the mass, and hence corresponded to a static
  bubble.} For $Q/R^2 >1$, there is no static bubble of any size, and
the black string was argued to decay to an expanding bubble.

Now let us return to the present case. Since our black string geometry
is not spherically symmetric, it is technically more difficult to
impose the condition that the geometry of the $S^3$ at the event
horizon agrees with the geometry of the $S^3$ where the circle
degenerates in the bubble. We will therefore content ourselves with a
rough estimate of the size. Of course, when the black string has an
AdS$_3 \times S^3$ near-horizon limit, this condition is trivially
satisfied, as the $S^3$ in the near-horizon region is a round $S^3$
with a radius determined by the charges. In the more general case, we
take $\sqrt{\tilde H_1 \tilde H_5}(r_+)$ as a rough estimate of the
size of the sphere at the horizon. for equal charges, this is
\begin{equation} \label{hbs} A \equiv \tilde H_1 (r_+) = r_+^2 + a_1^2
  \sin^2 \theta + M \sinh^2 \delta.
\end{equation}
Assuming $R \gg l_s$, \eqref{circle} implies $r_+^2$ will be small
compared to $\sqrt{\tilde H_1 \tilde H_5}$, and hence the first term
is negligible. Using \eqref{mc}, $a_1 = m M \sinh^2 \delta/R$, so 
\begin{equation}
  A = m^2 \frac{M^2}{R^2} \sinh^4 \delta \sin^2 \theta + M \sinh^2
  \delta = Q \left( 
    m^2 \frac{Q}{R^2} \tanh^2 \delta \sin^2 \theta + \tanh \delta
  \right).  
\end{equation}
We can make a similar estimate for the size of the minimum-area surface
$r=0$ in the soliton,
\begin{equation}
  \tilde A \equiv \tilde a_1^2 \sin^2 \theta + \tilde M \sinh^2 \tilde
  \delta = Q \left(
    m^2 \frac{Q}{R^2} \tanh^2 \tilde \delta \sin^2 \theta + \tanh \tilde
    \delta \right).
\end{equation}

There are then two cases: For $Q/R^2 \lesssim 1$, \eqref{soleq} and
\eqref{bseq} will give $\tanh \tilde \delta \sim 1$, $\tanh \delta
\sim 1$, and consequently $\tilde A \sim A$: the soliton is
roughly the same size as the black string. This includes the case $Q/R^2 \ll
1$, where the solution has an AdS$_3 \times S^3$ near-horizon
limit. 

For $Q/R^2 \gg 1$, \eqref{bseq} will give 
\begin{equation}
\tanh \delta \approx \left( \frac{R^2}{Q m^2} \right)^{1/3} \ll 1, 
\end{equation}
so 
\begin{equation}
A \sim Q \left( \frac{Q}{R^2} \right)^{1/3}. 
\end{equation}
For the soliton, there is now an important distinction between $m=1$
and $m \neq 1$. For $m=1$, $\tanh \tilde \delta =1$, and the soliton has
$\tilde A \sim Q^2/ R^2$, so 
\begin{equation}
\frac{\tilde A}{A} \sim \left( \frac{Q}{R^2} \right)^{2/3} \gg 1. 
\end{equation}
For $m \neq 1$, \eqref{soleq} gives 
\begin{equation}
\tanh \tilde \delta \approx \left( \frac{R^2}{Q (m^2-1)} \right)^{1/3}
\ll 1, 
\end{equation}
so 
\begin{equation}
\tilde A \sim Q \left( \frac{Q}{R^2} \right)^{1/3},
\end{equation}
and $\tilde A \sim A$. 

Thus, in almost all cases, the rough size of the minimum area surface
in the soliton and the horizon sphere in the black string agree. This
works even though both the black string and the soliton are far from
BPS in some cases.\footnote{The black string is always near-extreme,
  in the sense that it is close to the point where the horizon area
  vanishes, but it is far from the BPS bound if $\delta$ is small.}
The smooth soliton solution then provides a natural endstate for
tachyon condensation in the black string. The only exception is the
case $m=1$, $Q/R^2 \gg 1$, where the minimum radius in the soliton is
much bigger than the black string's horizon radius. It is not clear
what happens in this case. The soliton is still physically available
as a possible endstate for the tachyon decay in this case, but if the
black string decays by space pinching off locally at the black string
horizon, it will produce a bubble much smaller than the soliton. But
since the soliton is BPS in this case, it is a minimum-energy
configuration in the space of initial data, so it would be
energetically favorable for this bubble to evolve into the soliton.
There is still the alternative possibility that there is some other
endstate in this case: we have not made any attempt to identify such
an alternative endstate explicitly.  This seems a particularly
interesting issue for further study.

Some final issues concern the stability and uniqueness of the solitons
as endstates for the decay.  For $m=1$, the soliton is BPS, so it
clearly provides a stable endpoint for tachyon condensation for a
black string in which the circle along $\xi = \partial_y - \frac{1}{R}
\partial_\phi$ is string scale near the horizon. For larger $m$, the
situation is not clear, as the solitons of~\cite{nonbps} have not been
shown to be stable. However, we might view this as an indication that
they should be stable, so that they can provide an endpoint for the
tachyon condensation process. In the case $m=1$, more general smooth
solitons which break the rotational symmetry in the $\phi$ and $\psi$
directions have also been constructed~\cite{lm1,lm2,lmm}. It may be
possible to reach some of these other BPS solitons as the endpoint of
tachyon condensation by turning on the tachyon in a way which breaks
the symmetry, although to reach the general soliton solution, this
would have to produce a large change in the near-horizon geometry. In
the language of the worldsheet RG, there is a set of IR fixed points,
and which one we reach depends on the details of the trajectory.

We found no sign that expanding bubble solutions provide an endpoint
for tachyon condensation when we have periodic fermions on the
asymptotic $S^1$. The smooth stationary solitons provide a natural
endstate in almost all cases, and in the exceptional case, the soliton
is the state of lowest energy. This might have been anticipated, since
the positive energy theorem~\cite{witten:pose} forbids the existence
of analogues of the vacuum bubble of~\cite{witten:bubble} in this
case.  There is a family of vacuum bubble solutions where the circle
which shrinks is twisted, constructed by analytic continuation of the
Kerr metric~\cite{dggh,cg}; however, these solutions always have $|m|
<1$, so they do not have a compact circle asymptotically.\footnote{In
  the Euclidean Kerr black hole, the geometry has periodic
  identifications along the Killing vector which generates the
  horizons, $(\tau,\phi) \sim (\tau+2\pi R, \phi+ 2\pi R \Omega)$,
  where $\Omega$ is the angular velocity of the horizon and $R$ is the
  inverse of the surface gravity. It is found that $R \Omega <1$ for
  any choice of parameters, so there is no combination of this
  identification and $\phi \sim \phi + 2\pi$ which gives an
  identification simply along $\tau$, which would give a compact
  circle. See~\cite{dggh,cg} for details.} However, the standard
positive-energy results do not appear to forbid the existence of
finite-energy expanding bubbles, so this result is not simply a
consequence of those theorems. It is satisfying that we find no sign
that considering a localised finite-energy excitation will allow us to
initiate such a `decay to nothing' in the supersymmetric sector. This
is similar to results of~\cite{bath}, which found that bubbles
constructed from the Euclidean Kerr black hole do not expand into
regions with local supersymmetry.

That is, although we find that there are asymptotically supersymmetric
solutions with a localised tachyon, the decay of this tachyon does not
destroy the spacetime. Rather, the decay process is completely
localised (apart from the emission of some excess energy), and leaves
the asymptotics of the spacetime unchanged.  This is, in fact, an even
more localised example of tachyon condensation than the original
example of~\cite{panic}.  \smallskip

\textbf{Acknowledgements:} I thank Gary Horowitz for comments and
discussion. This work is supported by the EPSRC.

\bibliographystyle{/home/aplm/dma0sfr/tex_stuff/bibs/utphys}  
 
\bibliography{tachyons}

\providecommand{\href}[2]{#2}\begingroup\raggedright\begin{thebibliography}{10}

\bibitem{diss}
E.~J. Martinec, ``Defects, decay, and dissipated states,''
\href{http://xxx.lanl.gov/abs/hep-th/0210231}{{\tt hep-th/0210231}}.

\bibitem{over}
M.~Headrick, S.~Minwalla, and T.~Takayanagi, ``Closed string tachyon
  condensation: An overview,'' Class. Quant. Grav. {\bf 21} (2004)
  S1539--S1565,
\href{http://xxx.lanl.gov/abs/hep-th/0405064}{{\tt hep-th/0405064}}.

\bibitem{panic}
A.~Adams, J.~Polchinski, and E.~Silverstein, ``Don't panic! {C}losed string
  tachyons in {ALE} space-times,'' JHEP {\bf 10} (2001) 029,
\href{http://xxx.lanl.gov/abs/hep-th/0108075}{{\tt hep-th/0108075}}.

\bibitem{twist}
J.~R. David, M.~Gutperle, M.~Headrick, and S.~Minwalla, ``Closed string tachyon
  condensation on twisted circles,'' JHEP {\bf 02} (2002) 041,
\href{http://xxx.lanl.gov/abs/hep-th/0111212}{{\tt hep-th/0111212}}.

\bibitem{bk}
M.~Berkooz, Z.~Komargodski, D.~Reichmann, and V.~Shpitalnik, ``Flow of
  geometries and instantons on the null orbifold,''
\href{http://xxx.lanl.gov/abs/hep-th/0507067}{{\tt hep-th/0507067}}.

\bibitem{apart}
A.~Adams, X.~Liu, J.~McGreevy, A.~Saltman, and E.~Silverstein, ``Things fall
  apart: Topology change from winding tachyons,''
\href{http://xxx.lanl.gov/abs/hep-th/0502021}{{\tt hep-th/0502021}}.

\bibitem{hor:tachyon}
G.~T. Horowitz, ``Tachyon condensation and black strings,''
\href{http://xxx.lanl.gov/abs/hep-th/0506166}{{\tt hep-th/0506166}}.

\bibitem{chron}
M.~S. Costa, C.~A.~R. Herdeiro, J.~Penedones, and N.~Sousa, ``Hagedorn
  transition and chronology protection in string theory,''
\href{http://xxx.lanl.gov/abs/hep-th/0504102}{{\tt hep-th/0504102}}.

\bibitem{silv:tachyon}
J.~McGreevy and E.~Silverstein, ``The tachyon at the end of the universe,''
\href{http://xxx.lanl.gov/abs/hep-th/0506130}{{\tt hep-th/0506130}}.

\bibitem{witten:bubble}
E.~Witten, ``Instability of the {K}aluza-{K}lein vacuum,'' Nucl. Phys. {\bf
  B195} (1982)
481.

\bibitem{eva:handle}
A.~Saltman and E.~Silverstein, ``A new handle on de {S}itter
  compactifications,''
\href{http://xxx.lanl.gov/abs/hep-th/0411271}{{\tt hep-th/0411271}}.

\bibitem{gadsme}
V.~Balasubramanian, J.~de~Boer, E.~Keski-Vakkuri, and S.~F. Ross,
  ``Supersymmetric conical defects: Towards a string theoretic description of
  black hole formation,'' Phys. Rev. D {\bf 64} (2001) 064011,
\href{http://xxx.lanl.gov/abs/hep-th/0011217}{{\tt hep-th/0011217}}.

\bibitem{gadsmald}
J.~M. Maldacena and L.~Maoz, ``De-singularization by rotation,'' JHEP {\bf 12}
  (2002) 055,
\href{http://xxx.lanl.gov/abs/hep-th/0012025}{{\tt hep-th/0012025}}.

\bibitem{lunin}
O.~Lunin, ``Adding momentum to {D1-D5} system,'' JHEP {\bf 04} (2004) 054,
\href{http://xxx.lanl.gov/abs/hep-th/0404006}{{\tt hep-th/0404006}}.

\bibitem{gms1}
S.~Giusto, S.~D. Mathur, and A.~Saxena, ``Dual geometries for a set of 3-charge
  microstates,'' Nucl. Phys. {\bf B701} (2004) 357--379,
\href{http://xxx.lanl.gov/abs/hep-th/0405017}{{\tt hep-th/0405017}}.

\bibitem{gms2}
S.~Giusto, S.~D. Mathur, and A.~Saxena, ``3-charge geometries and their {CFT}
  duals,'' Nucl. Phys. {\bf B710} (2005) 425--463,
\href{http://xxx.lanl.gov/abs/hep-th/0406103}{{\tt hep-th/0406103}}.

\bibitem{nonbps}
V.~Jejjala, O.~Madden, S.~F. Ross, and G.~Titchener, ``Non-supersymmetric
  smooth geometries and {D1-D5-P} bound states,'' Phys. Rev. D {\bf 71} (2005)
  124030,
\href{http://xxx.lanl.gov/abs/hep-th/0504181}{{\tt hep-th/0504181}}.

\bibitem{bmpv}
J.~C. Breckenridge, R.~C. Myers, A.~W. Peet, and C.~Vafa, ``D-branes and
  spinning black holes,'' Phys. Lett. {\bf B391} (1997) 93--98,
\href{http://xxx.lanl.gov/abs/hep-th/9602065}{{\tt hep-th/9602065}}.

\bibitem{Gauntlett}
J.~P. Gauntlett, R.~C. Myers, and P.~K. Townsend, ``Black holes of {$D = 5$}
  supergravity,'' Class. Quant. Grav. {\bf 16} (1999) 1--21,
\href{http://xxx.lanl.gov/abs/hep-th/9810204}{{\tt hep-th/9810204}}.

\bibitem{herdeiro}
C.~A.~R. Herdeiro, ``Special properties of five dimensional {BPS} rotating
  black holes,'' Nucl. Phys. {\bf B582} (2000) 363--392,
\href{http://xxx.lanl.gov/abs/hep-th/0003063}{{\tt hep-th/0003063}}.

\bibitem{cyint}
M.~Cvetic and D.~Youm, ``Rotating intersecting {M}-branes,'' Nucl. Phys. {\bf
  B499} (1997) 253--282,
\href{http://xxx.lanl.gov/abs/hep-th/9612229}{{\tt hep-th/9612229}}.

\bibitem{klt}
P.~Kraus, F.~Larsen, and S.~P. Trivedi, ``The {C}oulomb branch of gauge theory
  from rotating branes,'' JHEP {\bf 03} (1999) 003,
\href{http://xxx.lanl.gov/abs/hep-th/9811120}{{\tt hep-th/9811120}}.

\bibitem{har}
T.~Harmark and N.~A. Obers, ``Thermodynamics of spinning branes and their dual
  field theories,'' JHEP {\bf 01} (2000) 008,
\href{http://xxx.lanl.gov/abs/hep-th/9910036}{{\tt hep-th/9910036}}.

\bibitem{lm2}
O.~Lunin and S.~D. Mathur, ``{AdS/CFT} duality and the black hole information
  paradox,'' Nucl. Phys. {\bf B623} (2002) 342--394,
\href{http://xxx.lanl.gov/abs/hep-th/0109154}{{\tt hep-th/0109154}}.

\bibitem{stat}
O.~Lunin and S.~D. Mathur, ``Statistical interpretation of {B}ekenstein entropy
  for systems with a stretched horizon,'' Phys. Rev. Lett. {\bf 88} (2002)
  211303,
\href{http://xxx.lanl.gov/abs/hep-th/0202072}{{\tt hep-th/0202072}}.

\bibitem{mathurrev}
S.~D. Mathur, ``The fuzzball proposal for black holes: An elementary review,''
\href{http://xxx.lanl.gov/abs/hep-th/0502050}{{\tt hep-th/0502050}}.

\bibitem{ls}
J.~M. Maldacena and H.~Ooguri, ``Strings in {AdS$_3$ and $SL(2,R)$ WZW model.
  I},'' J. Math. Phys. {\bf 42} (2001) 2929--2960,
\href{http://xxx.lanl.gov/abs/hep-th/0001053}{{\tt hep-th/0001053}}.

\bibitem{Friess}
J.~J. Friess, S.~S. Gubser, and I.~Mitra, ``Counter-examples to the correlated
  stability conjecture,''
\href{http://xxx.lanl.gov/abs/hep-th/0508220}{{\tt hep-th/0508220}}.

\bibitem{cvetic}
M.~Cvetic and D.~Youm, ``General rotating five dimensional black holes of
  toroidally compactified heterotic string,'' Nucl. Phys. {\bf B476} (1996)
  118--132,
\href{http://xxx.lanl.gov/abs/hep-th/9603100}{{\tt hep-th/9603100}}.

\bibitem{cvetlars}
M.~Cvetic and F.~Larsen, ``General rotating black holes in string theory:
  Greybody factors and event horizons,'' Phys. Rev. D {\bf 56} (1997)
  4994--5007,
\href{http://xxx.lanl.gov/abs/hep-th/9705192}{{\tt hep-th/9705192}}.

\bibitem{clads}
M.~Cvetic and F.~Larsen, ``Near horizon geometry of rotating black holes in
  five dimensions,'' Nucl. Phys. {\bf B531} (1998) 239--255,
\href{http://xxx.lanl.gov/abs/hep-th/9805097}{{\tt hep-th/9805097}}.

\bibitem{mm1}
E.~J. Martinec and W.~McElgin, ``String theory on {AdS} orbifolds,'' JHEP {\bf
  04} (2002) 029,
\href{http://xxx.lanl.gov/abs/hep-th/0106171}{{\tt hep-th/0106171}}.

\bibitem{mm2}
E.~J. Martinec and W.~McElgin, ``Exciting {AdS} orbifolds,'' JHEP {\bf 10}
  (2002) 050,
\href{http://xxx.lanl.gov/abs/hep-th/0206175}{{\tt hep-th/0206175}}.

\bibitem{lm1}
O.~Lunin and S.~D. Mathur, ``Metric of the multiply wound rotating string,''
  Nucl. Phys. {\bf B610} (2001) 49--76,
\href{http://xxx.lanl.gov/abs/hep-th/0105136}{{\tt hep-th/0105136}}.

\bibitem{lmm}
O.~Lunin, J.~Maldacena, and L.~Maoz, ``Gravity solutions for the {D1-D5} system
  with angular momentum,''
\href{http://xxx.lanl.gov/abs/hep-th/0212210}{{\tt hep-th/0212210}}.

\bibitem{witten:pose}
E.~Witten, ``A simple proof of the positive energy theorem,'' Commun. Math.
  Phys. {\bf 80} (1981)
381.

\bibitem{dggh}
F.~Dowker, J.~P. Gauntlett, G.~W. Gibbons, and G.~T. Horowitz, ``The decay of
  magnetic fields in {K}aluza-{K}lein theory,'' Phys. Rev. D {\bf 52} (1995)
  6929--6940,
\href{http://xxx.lanl.gov/abs/hep-th/9507143}{{\tt hep-th/9507143}}.

\bibitem{cg}
M.~S. Costa and M.~Gutperle, ``The {K}aluza-{K}lein {M}elvin solution in
  {M}-theory,'' JHEP {\bf 03} (2001) 027,
\href{http://xxx.lanl.gov/abs/hep-th/0012072}{{\tt hep-th/0012072}}.

\bibitem{bath}
O.~Aharony, M.~Fabinger, G.~T. Horowitz, and E.~Silverstein, ``Clean
  time-dependent string backgrounds from bubble baths,'' JHEP {\bf 07} (2002)
  007,
\href{http://xxx.lanl.gov/abs/hep-th/0204158}{{\tt hep-th/0204158}}.

\end{thebibliography}\endgroup

\end{document}